\documentclass[doublecol]{epl2}

\newcommand\Rm{{\rm Rm}}

\newcommand\Ru{{\rm Re}}
\newcommand\aver[1]{\langle#1\rangle}

\title{Long-term decaying evolution of MHD turbulence}
\shorttitle{Long-term decaying evolution of MHD turbulence} 

\author{P. Frick and R. Stepanov}
\shortauthor{P. Frick and R. Stepanov}

\institute{
  Institute of Continuous Media Mechanics, Korolyov str.1,
Perm, 614013, RUSSIA
}

\abstract{
The free decay of MHD turbulence at large Reynolds numbers is studied numerically using a shell model.
We study the statistical properties based on representative sample of realisations (128 realisations for each type of initial conditions) over the period of $10^5$ large-scale turnover times.
The performed simulations show that the force-free non-helical MHD turbulence can demonstrate two different scenarios of evolution in spite of similar initial conditions. Within the first scenario, the cross-helicity  accumulation is so fast that the energy cascade vanishes before significant magnetic energy dissipates. Then the system approaches the state of maximal cross-helicity.
Within the second scenario,  the cascade process continues to remain active until  time $10^4$ in units of large-scale turnover time. Then the  magnetic field becomes vastly helical due to magnetic helicity conservation. Thus the magnetic energy does not dissipate with kinetic energy.
}


\begin{document}

\maketitle

\section{Introduction}

Free decaying magnetohydrodynamic (MHD) turbulence
provokes interest for two main reasons. First, it raises the possibility of application to the physics of the interstellar medium and cosmology in the context of
evolution of the primordial magnetic field and its contribution to the present configuration of
the magnetic field in the Universe.
Second, the MHD turbulence differs from the conventional turbulence of incompressible fluids by an extended set of conservation laws, which form the basis of diverse scenarios of free evolution of turbulent motion.
This is a strong fundamental motivation for studying this problem.

Three ideal quadratic invariants are known in 3D incompressible magnetohydrodynamics: the total energy
$E=E^v+E^b$, the cross-helicity
$H^c=\aver{\vect{v}\cdot\vect{b}}$ and the magnetic helicity
$H^b=\aver{\vect{a}\cdot\vect{b}}$, where
$E^v=\aver{|\vect{v}|^2/2}$, $E^b=\aver{|\vect{b}|^2/2}$, $\vect{v}$ is the velocity filed, $\vect{b}$ is the magnetic field, $\vect{a}$ is the vector potential ($\vect{b}={\rm \nabla}\times\vect{a}$).
The cross-helicity characterises the correlation between the velocity
and magnetic field pulsations, and  magnetic helicity
characterises the correlation between the magnetic field and its
vector potential.
For the non-helical case ($H^b=H^c=0$), the decay law for the total energy of the {\it isotropic} MHD turbulence with equidistributed magnetic and kinetic energy ($E^b(k)\approx E^v(k)$) tends to $E(t)\sim t^{-1}$ \cite{Biskamp99}, which remains essentially slower than the decay of conventional turbulence, for which $E(t)\sim t^{-2}$.

The relation between the spectral density of magnetic helicity and the spectral density of magnetic energy
$|H^b(k)|\leq k^{-1}E^b(k)$ disables the direct spectral transfer of magnetic helicity to small scales \cite{Frisch75}. Thus the magnetic helicity condenses in the largest scale (smallest $k$) and holds there the corresponding part of magnetic energy.
Then the energy ratio $\Gamma=E^v/E^b$ goes down, and the spectral energy flux decreases along with the energy dissipation rate.
Biskamp and Muller \cite{Biskamp99} showed that, for finite magnetic helicity, the energy dissipation rate is governed by $H^b$ and $\Gamma$ as $\varepsilon=dE/dt\approx \Gamma^{1/2}(1+\Gamma)^{-3/2}E^{5/2}/H^b$. By making use of numerical result $\Gamma(t)\sim E(t)$ they  derived  for $\Gamma<<1$ an asymptotic law $E(t) \sim t^{-0.5}$. Campanelli \cite{Campanelli04} specified that
the kinetic energy decays as $E^v(t)\sim t^{-1}$ in both nonhelical and helical cases, while the magnetic energy follows the kinetic energy in the nonhelical case $E^b(t)\sim E^v(t)\sim t^{-1}$ and decays as $E^b(t)\sim  t^{-0.5}$ in the helical case independently of the initial conditions.
Direct numerical simulations \cite{Biskamp99,CHB} confirmed this phenomenology at relatively short time intervals (up to about 10 large-scale turnover time), available for computer modelling.

It is generally believed that for fully developed isotropic MHD turbulence  there are no reasons for
essential correlation between the pulsations of velocity and magnetic field (which means a noticeable level of cross-helicity).
 However, the discovery of highly correlated pulsation of velocity and magnetic field in the solar wind
\cite{solarwind} awoke interest in cross-helicity.
 As far back as 1980 the analysis of energy evolution  in free-decaying turbulence showed
that the cross-helicity decays more slowly than energy; therefore the correlation
between $\vect v$ and $\vect b$, characterised by the normalised cross-helicity
$C=H^c/E$, can grow over time  in free-decaying MHD turbulence \cite{dobrowolny80}.

The influence of cross-helicity on the forced MHD turbulence was
studied in \cite{gpl83}  in the context of the Alfvenic turbulence. Using EDQNM it was
shown that the system tends to a steady state, in which the
correlation coefficient is much higher than the ratio of helicity
to energy source injection rates. It has been also determined that the
energy spectrum of correlated MHD turbulence becomes steeper.
The role of cross-helicity in stationary forced  isotropic (not Alfvenic) MHD turbulence
has been studied by Mizeva et al. \cite{mizeva09}. It was shown that the injection of cross-helicity
suppresses the spectral energy transfer and leads to energy
accumulation in the turbulent flow. Then the spectrum becomes
steeper and the intermittency decreases. In the case of stationary forcing with uncontrolled injection of cross-helicity,  $C$ displays nontrivial behavior with long periods of high variability alternating with periods of almost constant $C$ \cite{2000EL.....52..539F}.

Turning back to decaying turbulence, let us note that cosmological applications require consideration of
very-long term evolution of MHD turbulence --- the age of the Universe calculated in  the unit of turnover time for the largest galactic turbulent scale ($\tau \approx 10^7$ years) gives $T \approx 10^4$, which is about 100 times longer than the best DNS range.
The required time series can be considered in the framework of shell models of turbulence.
Being an ultimate simplification of the original Navier-Stokes equations, shell models strictly provide the required
conservation laws, which are of special interest in the problems under discussion.
For the first time, the long-time evolution of free-decaying MHD turbulence has been considered by Antonov et al. \cite{Antonov01}.
Using the shell model of MHD turbulence introduced by Frick and Sokoloff \cite{FS98},
they performed  24 runs of long-time simulation (up to $t=10^4$ large-scale turnover time) with similar initial conditions. The developed Kolmogorov's  turbulence
with weak magnetic field was taken as an initial condition of the simulations. There was no special control of helicities. A coherent state with high alignment between the magnetic and
velocity fields, and essential reduction of the dissipation rate, were obtained for most realisations.
At the same time, there were a few realisations that displayed different behaviour,
characterised by a low level of cross-helicity.

In this paper we re-examine the long-time free decay  of MHD turbulence in the framework of shell models.
In contrast to previous work, we use a new (helical) shell model of MHD turbulence, introduced in \cite{mizeva09},
and we  study in detail the role of magnetic helicity and cross-helicity in the evolution scenario. We
use a massive computer cluster which allows us to consider the statistical properties based on the representative sample of realisations (128 realisations for each type of initial conditions).

\section{Shell model for helical MHD turbulence}

Shell models describe the dynamics of fully-developed MHD turbulence through a set of
complex variables $U_n$, $B_n$, which characterise the amplitudes of
velocity  and magnetic field pulsations in the shell of wave
number  $k_n<|\vect{k}|<k_{n+1}$, where $k_n=\lambda^n$ (
$\lambda$ is the shell width in a logarithmic scale). The total energy in terms of shell model is obviously
the sum of kinetic and magnetic energy of individual shells $E=\sum(|U_n|^2+|B_n|^2)/2$.
The cross-helicity is defined in a similar way as $H^c=\sum(U_nB_n^*+B_nU_n^*)/2$.
The definition of magnetic helicity is not so evident. The most popular MHD shell models
\cite{FS98,basu98} introduce the magnetic helicity as
$H^b=\sum (-1)^n |B_n|^2/k_n$, which associates the magnetic energy of a given shell with a
positive or negative magnetic helicity. Then the non-helical state can be obtained only through
the balance of magnetic energy in the even and odd shells.
Following  the idea used  for the hydrodynamical helicity in \cite{Melander},
we define the magnetic helicity as $H^b=\sum k_n^{-1}((B_n^*)^2-B_n^2)/2$. This definition allows us
to get helicity of any sign in any shell. So we use the model introduced in \cite{mizeva09}:
\begin{eqnarray}
 d_t U_n = i k_n (\Lambda_n(U,U)-\Lambda_n(B,B)) -
\frac{k_n^2 U_n}{\Ru} , \label{eq_su}\\ d_t B_n = i k_n
(\Lambda_n(U,B)-\Lambda_n(B,U)) - \frac{k_n^2 B_n}{\Rm},
\label{eq_sm}\end{eqnarray} where the nonlinear terms are written as
\begin{eqnarray}
\Lambda_n(X,Y)= \lambda^2 (X_{n+1}Y_{n+1}+X_{n+1}^*Y_{n+1}^*)
-X_{n-1}^r Y_n \nonumber \\ -X_n Y_{n-1}^r+\imath\lambda(2
X_n^*Y_{n-1}^i+X_{n+1}^r Y_{n+1}^i-X_{n+1}^i Y_{n+1}^r) \nonumber
\\
+X_{n-1}Y_{n-1}+X_{n-1}^*Y_{n-1}^* -\lambda^2(X_{n+1}^r Y_n
\nonumber +X_n Y_{n+1}^r) \\+\imath\lambda(2
X_n^*Y_{n+1}^i+X_{n-1}^r Y_{n-1}^i-X_{n-1}^i Y_{n-1}^r), \nonumber
\end{eqnarray}
a star means complex conjugation, and superscripts $r,i$ are real
and imaginary parts. $\Ru$ and $\Rm$ are the kinetic and magnetic
Reynolds numbers. In the limit $\Ru,\Rm \to \infty$ these eqs. (\ref{eq_su}) and (\ref{eq_sm}) conserve the
total  energy, the cross-helicity, and the magnetic helicity.
Time is measured in dimensionless units equal to the eddy turnover time
on the shell $n=0$ ($k_0=1$), which corresponds to the largest scale of the system.
For the case of hydrodynamics,  L'vov et al. \cite{1998PhRvE..58.1811L} suggested the parameter $\lambda$
equal to the golden number $(1+\sqrt{5})/2$ for optimal spectral resolution which is also successfully  applicable for MHD turbulence \cite{Stepanov06:JT}.

\section{Numerical results}

\begin{figure}
\includegraphics[width=0.49\textwidth]{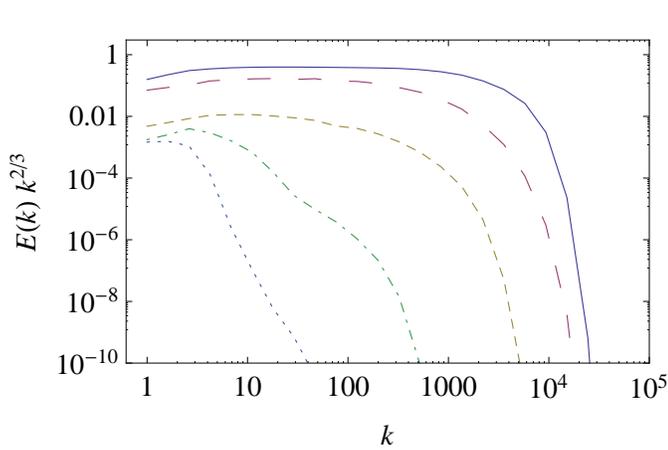}
\caption {Evolution of the compensated power spectrum, averaged on 128 realisations: $t=1, 10, 100, 10^3, 10^4$ (from top to bottom).}
\label{fig2}
\end{figure}

Equations (\ref{eq_su}) and (\ref{eq_sm})  were integrated for $0\leq n\leq40$ up to the time $t=10^5$, when any turbulent transfer has been finished and only pure exponential energy decay remains in the largest scales. In all simulations, $\lambda=1.618$ and $\Ru=\Rm =10^5$.
First we consider the decay of fully-developed turbulence with vanishing helicities $H^c$ and $H^b$.
Initial values of shell variables are $U_0=-\imath\sqrt{2}+\delta$, $B_0=\sqrt{2}+\delta$, where $\delta$ is a random complex additive with real and imaginary parts in a range $[-10^{-4}:10^{-4}]$, and  $U_n=B_n=0$  for other shells ($n>0$).
It corresponds to the initial state with $E^u\approx1$, $E^b\approx1$  and a small quantity of helicities $|H^c|\leq10^{-4}$,$|H^b|\leq10^{-4}$.

The inertial range with a Kolmogorov "-5/3"  power law is formed at about the turnover time of the vortex of maximal scale ($t\approx1$). In
fig.~\ref{fig2} we present the energy spectrum averaged on all 128 realisations at different stages of evolution. The spectrum is compensated by $k^{2/3}$ which leads to a horizontal line for the Kolmogorov spectral law. The
inertial range practically disappears at the time $t\approx10^3$, although the sagging in the middle part of the spectrum indicates that there are some realisations in which the energy is present at relatively small scales.

\begin{figure}
\includegraphics[width=0.49\textwidth]{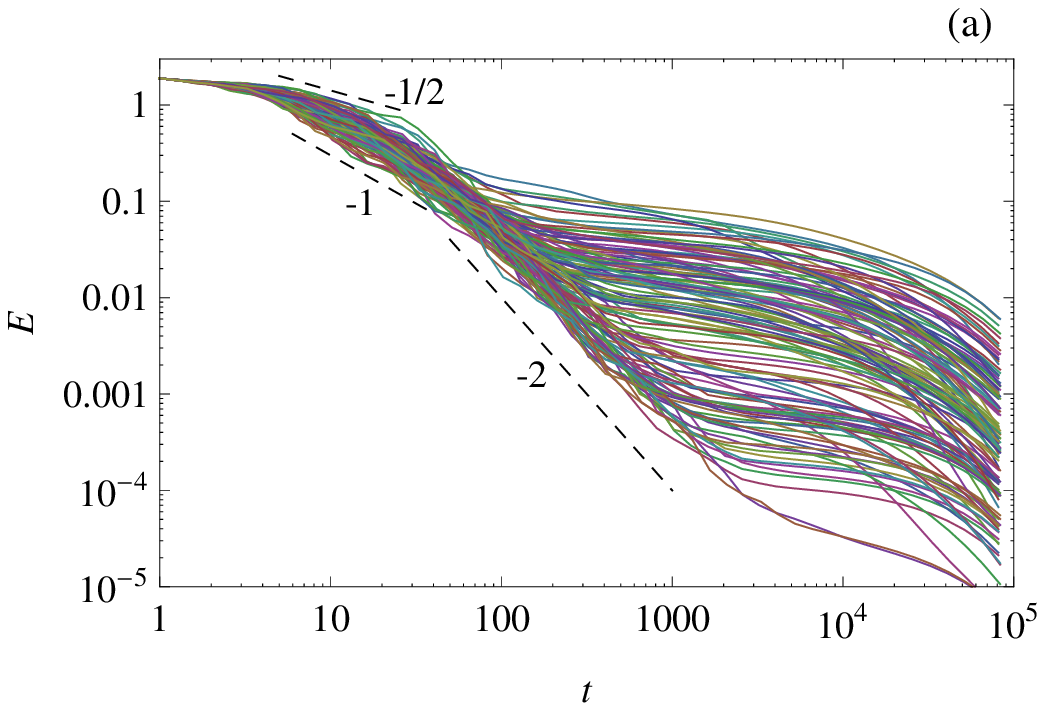}
\includegraphics[width=0.49\textwidth]{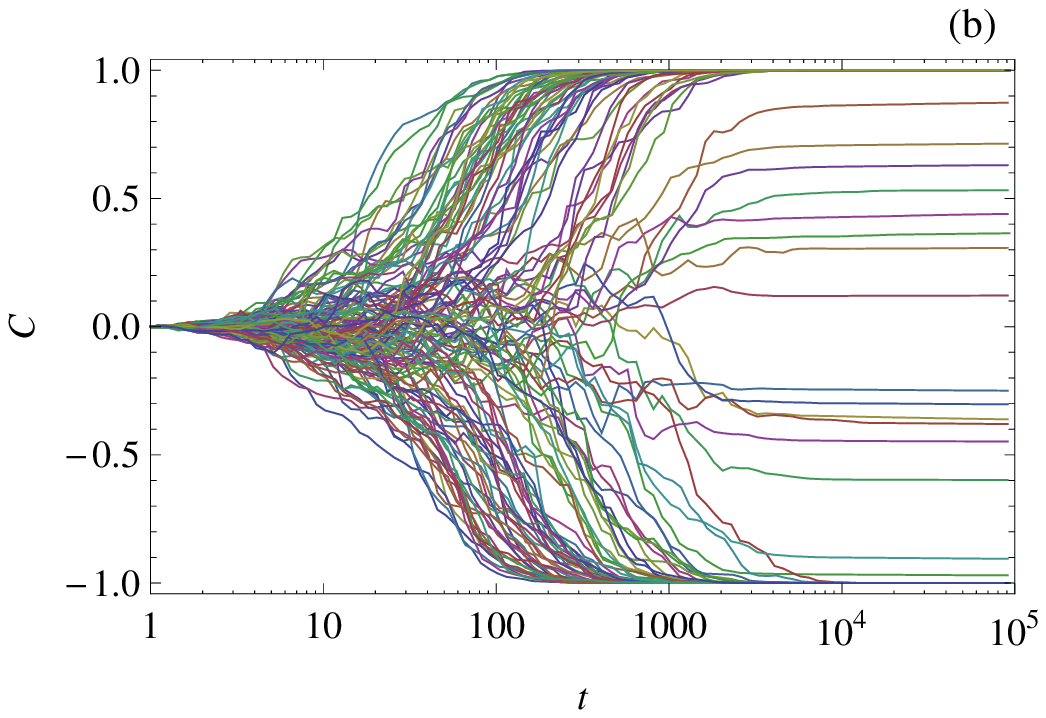}
\caption {Evolution of total energy (a) and normalised cross-helicity (b). 128 realisations with similar initial conditions.}\label{fig1}
\end{figure}

Figure~\ref{fig1}(a) shows the evolution of  the total energy in the entire totality of realisations. We see that similar initial conditions lead to the different scenarios in the evolution of the system. At the early stage of development (on the periods from several units to several ten) the bundle of trajectories remains sufficiently dense and is limited by the power laws $E(t)\sim t^{-1}$ and $E(t)\sim t^{-1/2}$, shown in the figure by dashed lines. Note that these are two power laws suggested for the decay of non-helical and helical MHD turbulence (in sense of magnetic helicity) \cite{Campanelli04}. At the next stage ($50< t < 1000$), the bundle of trajectories is bounded from below by a power law as before (but more steep, like $E(t)\sim t^{-2}$), while from above separate trajectories leave the bundle  practically horizontal, which indicates the vanishing cascade of energy and transition to the exponential dissipation of the energy.

The variety of scenarios of the evolution confirms fig.~\ref{fig1}(b), in which we show the evolution of the normalised cross-helicity $C$. The quantity $C=H^c/E$ characterizes the part of energy, concentrated in the  correlated pulsations of velocity and magnetic field. The limit $C \to \pm 1$ corresponds to a completely correlated state ($U_n=B_n$), in which the nonlinear energy transfer is blocked. Figure~\ref{fig1}(b) shows that the main part of the trajectories reaches this state in the range of time $100< t<1000$, but there are some trajectories (about 10--15\%) for which the evolution of $C$ stops at some arbitrary level.

\begin{figure}
\includegraphics[width=0.49\textwidth]{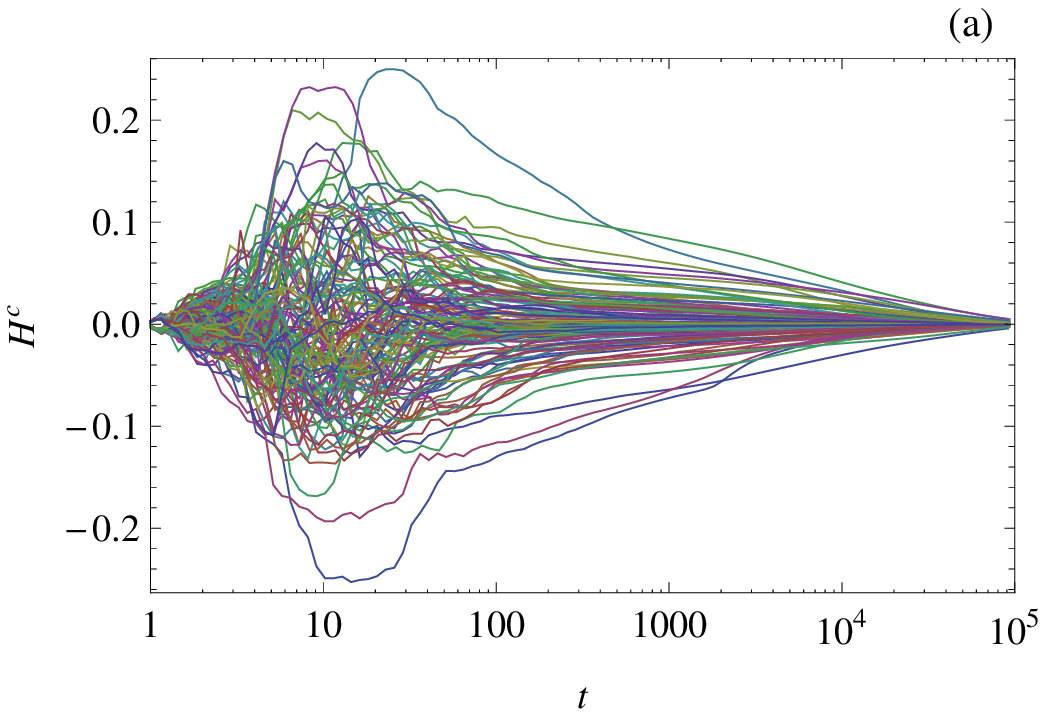}
\includegraphics[width=0.49\textwidth]{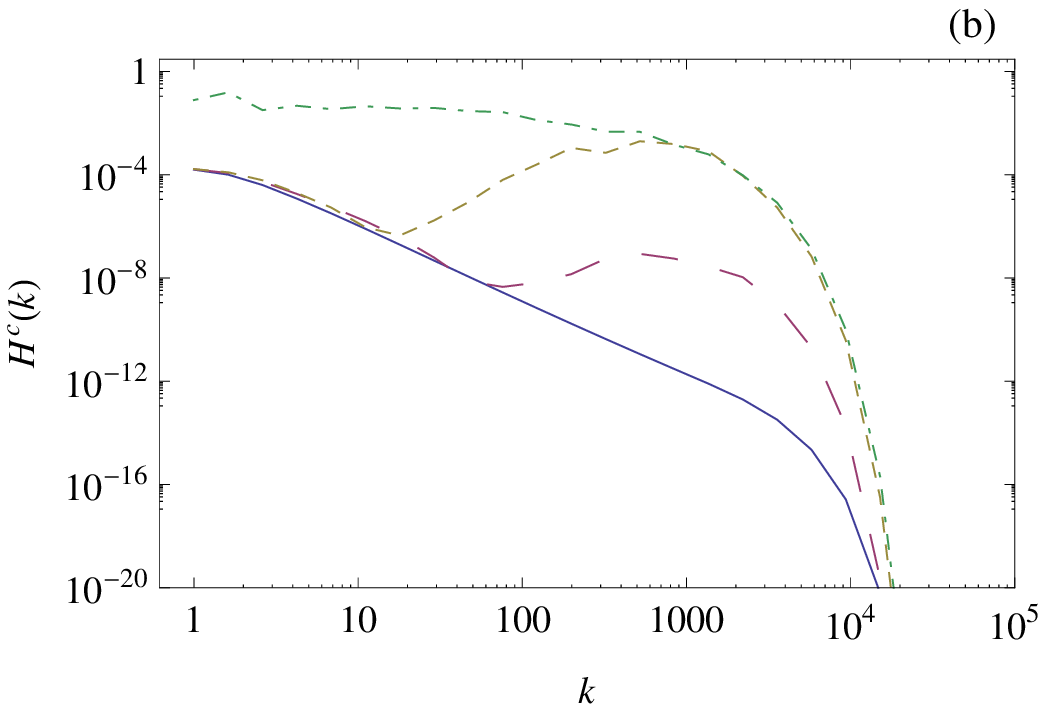}
\caption {Evolution of  the cross-helicity $H^c$  for all realisations (a) and spectral distribution for a single realisation at time t=0.8, 0.9, 1, 5 (lines from bottom to top) (b).   }
\label{fig3}
\end{figure}

Figure~\ref{fig3}(a) shows that the cross-helicity $H^c$ can be generated, in contrast to energy, which can decay only.
The active cross-helicity production is mostly observed until $t\approx10^2$. In some realisations the cross-helicity reaches a level $|H^c|\approx 0.2$. The source of the cross-helicity in the force-free evolution can be the dissipation term only.  In fig.~\ref{fig3}(b) we present  the evolution of the spectrum of cross-helicity for one realisation (namely, we took the realisation  that corresponds to the lowest trajectory in fig.~\ref{fig3}(a)). One can see  that most intensive production of the $H^c$ happened at the dissipation scale. Then nonlinear terms transfer $H^c$ to the largest scale like an { \it inverse cascade}. Note that in contrast to the curves in fig.~\ref{fig2}, the lower curves in fig.~\ref{fig3}(b) correspond to earlier moments of time.

The cross-helicity, which can be produced in the smallest scales by dissipation, is transported towards large scales through the inertial range. Then the spectral energy flux is  considerably reduced depending on the level of accumulated cross-helicity in a given realisation. This leads to the blocking of the turbulent energy cascade, which occurred in different realisations at substantially different moments in time and with  substantially different values of the remaining total energy (see fig.~\ref{fig1}a).
However, the growth of the normalised cross-helicity does not result in the completely correlated state $C=\pm1$ for all realisations (see fig.~\ref{fig1}b). Some realisations continue to stay at values $-1<C<1$.
These realisations develop a highly helical magnetic field. The magnetic helicity (which is weak at the initial state) does not cascade to small scales  and practically does not dissipate. This has been demonstrated through various studies starting from \cite{Frisch75}. If the energy transfer (and dissipation) is not blocked by the cross-helicity, only the magnetic field with maximal helicity survives at the late stage of the evolution.

\begin{figure}
\includegraphics[width=0.45\textwidth]{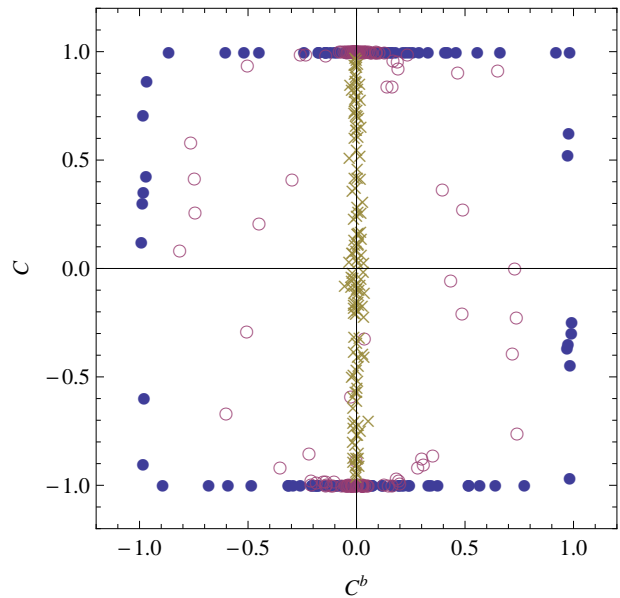}
\caption {normalised cross-helicity  $C$ vs normalised magnetic helicity $C^b$ for different realisations at $t=10^2$ (crosses), $t=10^3$ (open circles) and $t=10^4$ (full circles).}
\label{fig4}
\end{figure}
\begin{figure}
\includegraphics[width=0.45\textwidth]{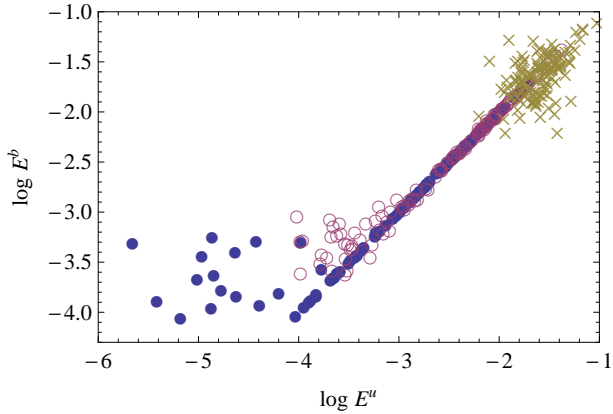}
\caption {Magnetic energy  $E^b$ vs kinetic energy $E^u$ for different realisations at $t=10^2$ (crosses), $t=10^3$ (open circles) and $t=10^4$ (full circles).}
\label{fig6}
\end{figure}

The tendencies of correlation evolution  are illustrated fig.~\ref{fig4}, in which the distribution of realisations is shown on the ($C^b$,$C$) plane at different moments in time (here $C^b=H^b/(k_0 E^b)$ ).
All realisations start from the origin of the graph (at $t=0$).
At the first stage (up to $t\approx100$), points scatter along the vertical line, showing the
rapid increase of $|C|$. Several realisations ($\approx 10\%$) deviate from the axis $C^b=0$. At the time $t=10^3$, this set of points forms a cloud in the center of the plane, while the rest are concentrated on the lines $C=\pm1$. At the late stage ($t=10^4$), practically all points are on the lines $C=\pm1$ or $C^b=\pm1$.
The first case ($C=\pm1$) means that magnetic and velocity fields are completely correlated, while the second case
($C^b=\pm1$) means that only the helical magnetic field remains at the largest scale.

Figure~\ref{fig6} shows the relation between the kinetic and magnetic energies for all runs at different times. One can see that both energies scatter randomly from $0.01$ to $0.1$ at early stage ($t=100$). At time $t=10^3$, kinetic and magnetic fields approach the equipartition state ($U_n=B_n$) and the points mainly lie close to the line $E^b=E^u$. At late stage ($t=10^4$) some points move away from this line. These points correspond to  realisations that follow the second scenario,  in which the kinetic energy continues to decay while the magnetic one, being helical,  does not change  ($C^b=\pm1$).

\section{Discussion and Conclusions}

The performed simulations show that the force-free non-helical ($H^b\approx0, H^c\approx0$) MHD turbulence can demonstrate fundamentally different ways of evolution in spite of similar initial conditions. We distinguish two scenarios of evolution. Within the first scenario, the cross-helicity accumulation is so fast that the energy cascade vanishes before significant magnetic energy dissipates. Then the system comes to a state with $C=\pm1$ and the value of $C^b$ depends on the rest of the magnetic energy.
Within the second scenario,  the cascade process remains active until the late time $t\sim10^4$, when the  magnetic field becomes vastly helical and later magnetic energy does not dissipate with kinetic energy.
Then the system comes to the final state with $C^b=\pm1$ and an arbitrary value of $C$.

Comparison of the results with a direct numerical simulations is hardly possible, because in high resolution DNS the  reachable time is typically less than 100 dimensionless units. Shell model simulations have shown that the difference in evolution scenario is well pronounced after $10^3$ units of time (see fig.~\ref{fig6}).

The probability of appearance of both scenarios depends on the presence of helicities $H^c$ and $H^b$ in the initial distribution. We performed simulations with initial conditions $E^u(k_0)=E^b(k_0)=1$ and some given value of $C$ or $C^b$ in all 128 realisations. It was found that
the presence of even a small portion of magnetic helicity substantially increases the probability of the second scenario: $C^b=0.01$ at $t=0$ leads  the second scenario being developed in half of the cases, and initial value $C^b=0.05$ leads to the likelihood that practically all realisations will follow the second scenario.
The presence of cross-helicity at the beginning of evolution is not so crucial (because the turbulence itself generates the cross-helicity). $C=0.01$ at $t=0$ does not change the distribution of trajectories at all. Only $C\approx 0.1$ is enough to exclude the appearance of the second scenario.

We stress the point that the dynamics of helicities has a signifiable influence on  the evolution of free-decaying MHD turbulence.     Our understanding is based on the idea of the inverse cascade of magnetic helicity \cite{Frisch75} which has been confirmed by numerous simulation \cite{1996PhRvD..54.1291B,CHB,2001ApJ...550..824B,2005PhRvE..71d6304M,2006ApJ...640..335A}. Alexakis et al \cite{2006ApJ...640..335A} reported that smaller-in-amplitude direct cascade is observed from the largest scale to small scales. This may be a result of insufficient inertial range resolution, in which forcing and dissipation scales are not well separated. In shell models, the Reynolds number is always quite large ($10^5$ or more) and the direct cascade of magnetic helicity was never observed.
From the other perspective, a shortcoming of shell models is ignoring the nonlocal (in scales) interactions, due to which  weak direct cascade of magnetic helicity may occur. This shortcoming can be overcome in the framework of shell models -- a detailed study of the shell-to-shell interactions \cite{2005PhPl...12d2309D} gives a  base on which to build
a  nonlocal MHD shell model \cite{2007NJPh....9..294P}.

\begin{figure}
\includegraphics[width=0.40\textwidth]{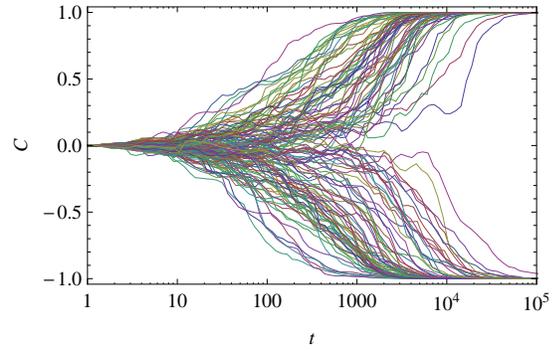}
\caption {Evolution of normalised cross-helicity in a sample of 128 realisations in the system with extended large-scale spectral range ($k_{min}=0.1$).}\label{fig5}
\end{figure}

Finally, we note that the relationship $|H^b(k)|\leq k^{-1}E^b(k)$ indicates that the character of evolution can be changed substantially if the magnetic helicity can move to scales larger than the scale of maximal energy at $t=0$.
To examine this case, we extended the range of scales into the red part of the spectrum, simulating eqs.~(\ref{eq_su}) and (\ref{eq_sm}) for $-5\leq n\leq 35$. This means that the maximal scale accessible for the turbulence becomes about 10 times larger than the scale in which the energy is set at the initial state ($k_{min}\approx 0.1k_0$). The result of the  simulation is shown in fig.~\ref{fig5}. In contrast to fig.~\ref{fig1}(b), there are no more realisations that avoid the completely correlated final state ($C=\pm 1$). The inverse cascade of magnetic helicity leads to the fact that an increasingly smaller part of magnetic energy might be blocked at the largest scales, being excluded from the direct energy cascade
to small scales. Then only the correlated part of velocity and magnetic fields survives at the late state of evolution.

\acknowledgments
This work was supported by ISTC  (project 3726) and Russian Academy of Science (projects 09-$\Pi$-1-1002).
The support of the parallel computations on the supercomputer
SFIF MSU "Chebyshev" is kindly appreciated.

\bibliographystyle{eplbib}
\bibliography{ref}

\end{document}